\documentclass[12pt]{article}
\usepackage{graphicx}
\usepackage{times}

\usepackage{float}

\usepackage{fancyhdr}
\usepackage{epsfig}
\usepackage{multirow}
\usepackage{color}
\usepackage{amsfonts,amsmath}

\begin{document}
\title{Optimizing surveillance for livestock disease spreading through animal movements}

 \author{Paolo Bajardi${}^{1,2}$, Alain Barrat${}^{2,3}$, Lara Savini${}^{4}$, Vittoria Colizza${}^{5,6,7}$}

\date{}
\maketitle

\begin{abstract}
{ 
\noindent
The spatial propagation of many livestock infectious diseases critically depends on the animal movements among premises, so that the knowledge of movements data may help to detect, manage and control an outbreak. The identification of robust spreading features of the system is however hampered by the temporal dimension characterizing population interactions through movements. Traditional centrality measures do not provide relevant information as results strongly fluctuate in time and outbreak properties heavily depend on geotemporal initial conditions. By focusing on the case study of cattle displacements in Italy, we aim at characterizing livestock epidemics in terms of robust features useful for planning and control, to deal with temporal fluctuations, sensitivity to initial conditions, and missing information during an outbreak. Through spatial disease simulations, we detect spreading paths that are stable across different initial conditions, allowing the clustering of the seeds and reducing the epidemic variability. Paths also allow us to identify premises, called sentinels, having a large probability of being infected and providing critical information on the outbreak origin, as encoded in the clusters. This novel procedure provides a general framework that can be applied to specific diseases, for aiding risk assessment analysis and informing the design of optimal surveillance systems.

  }

\end{abstract}

\footnotetext[1] {Computational Epidemiology Laboratory, Institute for Scientific Interchange (ISI), Turin, Italy}
\footnotetext[2]{Centre de Physique Th\'eorique, Aix-Marseille Univ, CNRS UMR 6207, Univ Sud Toulon Var, 13288 Marseille cedex 9, France}
\footnotetext[3]{Data Science Laboratory, Institute for Scientific Interchange (ISI), Turin, Italy}
\footnotetext[4]{Istituto Zooprofilattico Sperimentale Abruzzo-Molise G. Caporale, Teramo, Italy}
   \footnotetext[5]{INSERM, U707, Paris F-75012, France}
   \footnotetext[6]{UPMC Universit\'e Paris 06, Facult\'e de M\'edecine Pierre et Marie Curie, UMR S 707, Paris F75012, France}
 \footnotetext[7] {Institute for Scientific Interchange (ISI), Turin, Italy}

\section{INTRODUCTION}

Livestock infectious diseases represent a major concern as they may
compromise livestock welfare and reduce productivity, induce large
costs for their control and eradication~\cite{Anderson:2002}, and may
in addition represent a threat to human health, since the emergence of
human diseases is dominated by zoonotic
pathogens~\cite{Taylor:2001}. Disease management and control are thus
very important in order to reduce such risks and prevent large
economical
losses~\cite{Ferguson:2001,Keeling:2001,Kao:2003,Keeling:2005,Carpenter:2011},
and strongly depend on our ability to rapidly and accurately detect an
outbreak and protect vulnerable elements of the system. The major
difficulty lies in the assessment and prediction of the potential
consequences of an outbreak, and how these depend on specific conditions
of the epidemic event. Control may be hampered by the non-localized nature
of disease transmission, with animal movements facilitating the
geographical spread of the diseases on large spatial scales~\cite{Anderson:2002}. 
The knowledge of the pattern
of movements among populations of hosts is thus crucial in that it
represents the key driver of infection spread, defining the substrate
along which transmission can occur.  The availability of detailed
datasets of animal movements allows for the explicit analysis of these 
patterns and the simulation of the spatial spreading  of animal
diseases among premises, aimed at the characterization of premises in
terms of their risk of exposure or spreading
potential~\cite{Christley:2005,Ortiz:2006,Bigras:2006,Green:2006,Kao:2006,Robinson:2007,Kao:2007,Natale:2009,Dube:2009,Martinez:2009,Vernon:2009,Volkova:2010,Rautureau:2010,Natale:2011,Bajardi:2011}.

A network representation~\cite{Barabasi:1999,Dorogovstev:2003,Newman:2003,BBVbook}
is a natural description of the set of animal movements with 
nodes corresponding to  livestock-holding locations, and links
referring to  livestock movements. Network approaches to epidemic
spreading are widely used, leading to valuable and important results
in the understanding of the system's properties relevant to the disease spreading.  
Different centrality measures have been investigated in
order to identify the nodes with largest spreading potential that
should be targeted for disease control~\cite{BBVbook,Friedkin:1991,Albert:2000,Cohen:2001,Pastor:2001,Lloyd:2001,Kitsak:2010}, with a focus mainly  on the static properties of
the spatial and topological aspects of contact and movement patterns.
The study of livestock trade movement data, however, has shown the presence of large heterogeneities
characterizing the network from the geographical and temporal point of
view~\cite{Christley:2005,Ortiz:2006,Bigras:2006,Kao:2006,Robinson:2007,Natale:2009,Volkova:2010,Rautureau:2010,Woolhouse:2005,Brennan:2008}, and of a strong dynamical activity at the local level that limits the usefulness of
projections into static properties~\cite{Bajardi:2011}. The temporal nature of the
pattern of livestock movements thus opens novel challenges limiting
our understanding of the epidemic process because of (i) the strong
dependence of the spreading pattern on the initial conditions, both
geographical and temporal~\cite{Green:2006}, and (ii) the lack of
meaningful definitions of nodes' importance, given the observed large
temporal fluctuations of centrality measures based on static
structural properties~\cite{Bajardi:2011}. Both aspects limit
our ability to design robust and efficient surveillance and containment measures
by strongly increasing the number of degrees of freedom 
  responsible for the outbreak outcomes.

Here we address these
challenges by considering the spread of livestock diseases on the
dataset of cattle displacements among Italian animal
holdings~\cite{Natale:2009,Bajardi:2011}, where the full temporal
resolution of the dataset is considered.  
In order to gain a general
understanding of the interplay between the spreading dynamics and the
temporal features of the animal movements, we consider a simple model
of a notifiable highly contagious disease characterized by short
timescales where the single epidemiological unit corresponds to the
farm (i.e. the node of the network) and transmission can occur from
farm to farm through animal movements (i.e. the
links of the network)~\cite{Keeling:2005}.  
We propose a novel method, applied to the
dataset under study, that uncovers the presence of similar spreading patterns
allowing the clustering of initial conditions, thus reducing the number of degrees of freedom,
and the identification of sentinel nodes to be targeted
for disease surveillance.
  Appropriately
parameterized applications can be considered for specific livestock
diseases where movement-related transmission is a considerable risk
factor.

%*******************************************************************************************************
\section{MATERIALS AND METHODS}

%************************************************
\subsection{Dataset and network representation}

\begin{figure}[H]
\centerline{\includegraphics[width=0.9\textwidth]{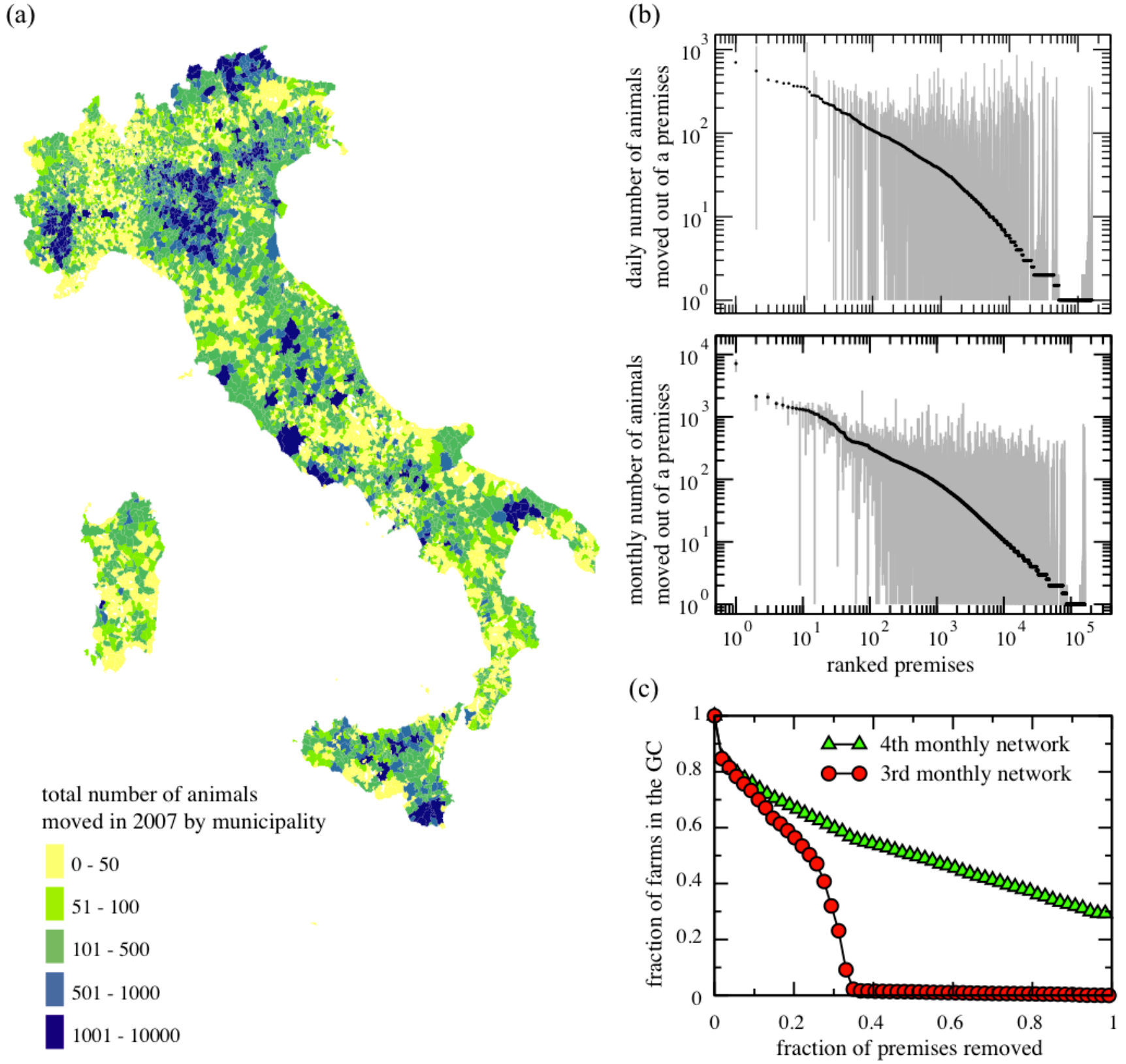}}
\caption{Properties of the cattle movement dataset. (a) Geographical
  representation of the total number of animals moved during the year
  2007 for each municipality of the country. The color code is
  assigned according to the outgoing fluxes of displaced bovines. (b)
  Median (black) and $95\%$ confidence intervals (grey) of outgoing
  traffic of each premises. For the sake of visualization the premises
  have been ranked by the median values. The traffic has been evaluated
  on daily (top) and monthly basis (bottom). (c) Two
  consecutive monthly networks ($n=3$ and $n=4$) have been
  considered. A list of premises with decreasing number of connections is
  calculated on the snapshot $n=3$, and is applied as a removal
  strategy for both networks, i.e. from best connected premises to least connected
  ones, calculated on the snapshot $n=3$ only. The relative size of the giant component GC
  (i.e. the largest fraction of premises that are connected with each other
  and thus potentially reachable by the disease) is shown as a function of the
  fraction of premises removed.}
\label{fig:dataset}
\end{figure}

The data on cattle trade movements used in the present study is obtained from the 
Italian National Bovine database
and provides a daily description of the movements of each 
bovine in Italy, specifying the premises of origin and destination and the date 
of the movement for each animal (identified through a unique ID)~\cite{Natale:2009}. The dataset refers 
to the year $2007$ and contains the
movements of almost $5$ million bovines between more than $170,000$
premises involving $96\%$ of the Italian municipalities (see Figure~\ref{fig:dataset}a)~\cite{Natale:2009}. 
The dataset can be described through a
dynamical network~\cite{Green:2006,Vernon:2009,Bajardi:2011,Keeling:2010} where the nodes correspond to premises and a directed link
represents a displacement of bovines between two premises. 

By aggregating all the displacements that take place within a given time interval
$[n\Delta t,(n+1)\Delta t]$, it is possible to construct a series of temporally
ordered static networks describing the movements at a temporal
resolution $\Delta t$. The $365$ daily networks ($\Delta
t=1$) correspond to the finest available temporal resolution, but other time scales
(such as $\Delta t=7$, $\Delta t=28$, $\Delta t=365$) may be used~\cite{Christley:2005,Ortiz:2006,Bigras:2006,Kao:2006,Natale:2009,Volkova:2010,Rautureau:2010,Woolhouse:2005}.

%*******************************************************************************************************
\subsection{Epidemic simulations on the dynamical network of cattle movements}

The disease spread on the dynamical network is modeled using a simple
SIR compartmental model~\cite{Anderson:1992}.  We assume that premises
 are the discrete single units of the process, neglecting the
possible impact of within-farm dynamics, as commonly assumed in the
study of the spread of highly contagious and rapid infectious diseases
through animal movements~\cite{Keeling:2005}. Premises are labeled as
Susceptible, Infectious, or Removed, according to the stage of the
disease. All premises are
considered susceptible at the beginning of the simulations, except for
the single seeding farm. At each time step, an infectious farm $i$ can
transmit the disease along its outgoing links to its neighboring
susceptible farms that become infected and can then propagate the
disease further in the network.  Here we consider a deterministic
process for which the contagion occurs with probability equal to 1 as
long as there is a directed link of cattle movements from an
infectious farm to a susceptible one at a given time step~\cite{Green:2006}. Though
a crude assumption, this allows us to simplify the computational
exploration of the initial conditions, focusing on the fastest
infection patterns.
The corresponding stochastic case is reported in the
Electronic Supplementary Material (ESM), where both high and intermediate
transmissibility rates are considered.
After $\mu^{-1}$ time steps an
infected farm becomes recovered and cannot be reinfected. The
simulation is fully defined by the choice of the timescale $\Delta t$,
used to define the successive aggregated networks 
 and of the initial conditions $(x_0,t_0)$ where
$x_0$ is the seeding node  and $t_0$ indicates the outbreak start.

%************************************************
\subsection{Invasion paths and seeds' cluster detection}

Given the limited applicability of quantities defined {\em a priori} to characterize the 
spreading potential of a node in such a highly dynamical network, here we exhaustively 
explore the dependence of the spreading process on the initial conditions and investigate
the possible emergence of recurrent patterns, aiming at identifying similar spreaders
in such a complex environment. The disease spreading pattern  is
 encoded in an invasion path characterized by 
 a set of nodes $\vec\nu$, a set of directed links $\vec l$ (indicating the transmission), and a seed $x_0$.  
 We define the overlap
$\Theta_{12}$ between two paths $\Gamma_1$ and $\Gamma_2$ as the
Jaccard index
$\frac{|\vec\nu_1\bigcap\vec\nu_2|}{|\vec\nu_1\bigcup\vec\nu_2|}$,
measuring the number of common nodes over the total number
of nodes reached by the two paths. This
measure does not consider the information on the links of transmission
from one farm to another, as we are interested in the observable
outcome of the outbreak, namely the fact that a farm is infected or
not, rather than  the precise transmission path. We also tested an
alternative definition of the overlap taking
into account the directed links composing the paths (see the ESM).

\begin{figure}[H]	
\begin{center}
\centerline{\includegraphics[width=0.9\textwidth]{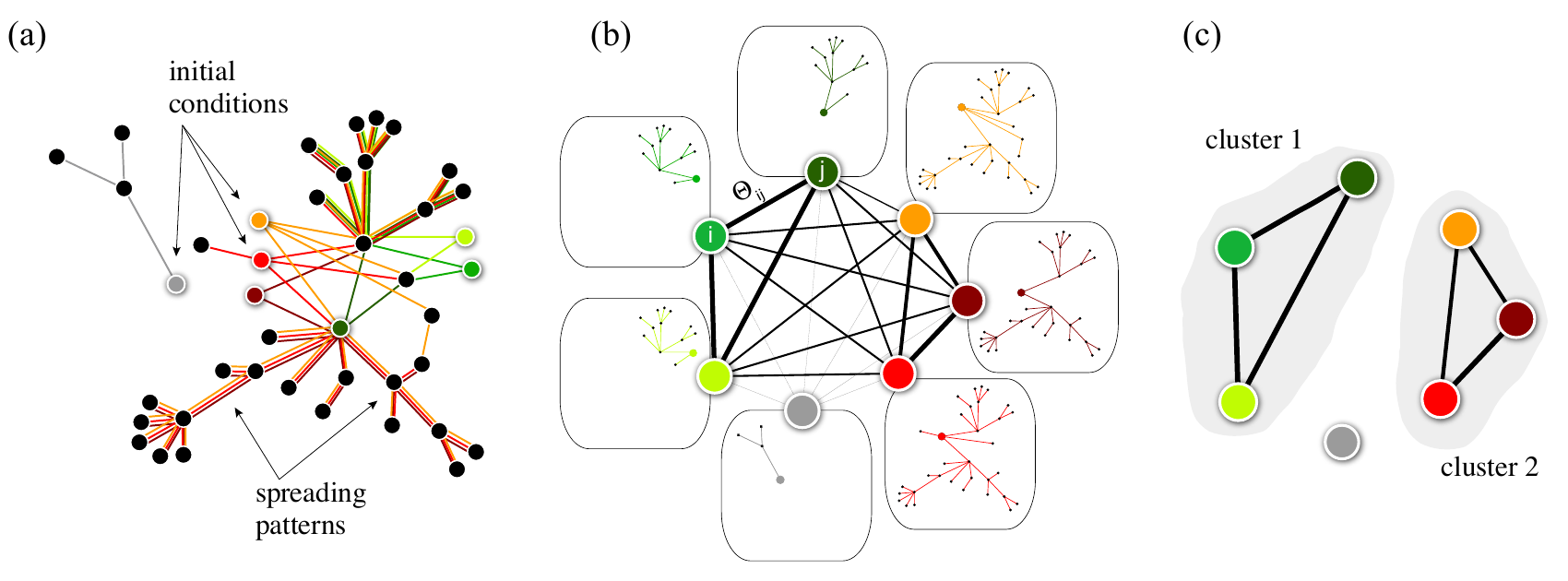}}
\caption{Schematic representation of the cluster detection
  procedure. (a) Different simulated invasion paths (colored lines)
  obtained for different seeder (corresponding colored nodes) are
  shown on the network. (b) The initial conditions similarity network
  (ICSN) is obtained by calculating, for any pair of initial
  conditions $i$ and $j$, the overlap $\Theta_{ij}$ measuring the
  similarity between the invasion paths originated by $i$ and
  $j$. Thicker lines in the ICSN indicate a higher overlap. (c) By
  removing all links of the ICSN with an overlap lower than a given
  threshold $\Theta_{th}$, clusters of nodes leading to similar
  propagation paths emerge.
  }
\label{fig:clustering}
\end{center}
\end{figure}

We have computed, at fixed $\Delta t=1$ day and initial time
$t_0$, the overlap $\Theta_{12}$ between the invasion paths of
deterministic SIR outbreaks generated by every pair of potential seeds
$(x_1,x_2)$ and  constructed the initial conditions similarity
network (ICSN)
as a weighted undirected network in which each node is
an initial condition of the epidemic  and the link between two
nodes $x_1$ and $x_2$ is weighted by the value of the overlap
$\Theta_{12}$, measuring the similarity of the invasion paths they
produce.  By filtering the ICSN to disregard values of the similarity
smaller than a given threshold $\Theta_{th}$, 
subsets of nodes with similar spreading properties may emerge (see 
Figure~\ref{fig:clustering}).

The method described above unveils a partition ${\cal P}(t_0)$ of the 
possible seeds that  depends
on the starting time $t_0$ of the spreading.
In order to measure the robustness of the clusters $C_i(t_0)$ at time $t$, 
we define the vector $\vec\rho_i(t,t_0)$ with components
$\rho_{i,j}(t,t_0)=\frac{|C_j(t)\bigcap C_i(t_0)|}{|C_i(t_0)|}$,
representing the fraction of nodes of $C_i(t_0)$ present in the
cluster $C_j(t)$. 
If the partitions are equal
at times $t_0$ and $t$, each vector $\vec\rho_i(t,t_0)$ will have one
component equal to $1$, and all the others equal to $0$. If instead
the nodes of $C_i(t_0)$ are homogeneously redistributed into
the $C$ clusters $C_j(t)$ of ${\cal P}(t)$, $\vec\rho_i(t,t_0)$ will have all components
equal to $1/C$.
Here we consider the $C=20$ largest clusters for each $t$ and measure 
the conditional entropy
\begin{equation}
H_i(t,t_0)=\frac{1}{\sigma_i(t,t_0) \log(\sigma_i(t,t_0) / C)} \sum_j \rho_{i,j}(t,t_0)\log\rho_{i,j}(t,t_0)\,
\end{equation} 
of observing a specific redistribution
among the largest $C$ 
clusters at time $t$, given that only a fraction $\sigma_i(t,t_0)=\sum_j \rho_{ij}(t,t_0)$
of the
original nodes are found within those clusters, rescaled by its
maximum value.
If $C_i(t_0)$ is also a cluster of
${\cal P}(t)$, $H_i(t)=0$. If its nodes are equally divided into
the $C$ clusters of ${\cal P}(t)$, the entropy is equal to $1$.  In
general the entropy takes values in the interval 
$\left[ (1-\log C/\log\sigma_i)^{-1}, 1\right]$, 
where its minimum value, $\min_{H_i(t)}$, represents
the best possible configuration; all the nodes of $C_i(t_0)$ are in the same
cluster of ${\cal P}(t)$, except the fraction $(1-\sigma_i)$ that do
not belong anymore to the largest $C$ clusters.  
We explore in the ESM additional quantities to measure the stability of the
partitions. 

%************************************************
\subsection{Uncertainty in the identification of the seed cluster}
The presence of similar invasion paths may be exploited for the identification of the
seed cluster starting from a specific infected premises. In order to investigate whether
this is possible, we explore all paths of infections and  measure the number of times that any node  in the network is 
reached by the epidemic,  breaking down this number according to the seed cluster
originating the epidemic. We  then associate to each node $k$, reached by the disease $n_k$ times, a
vector $\vec{\pi} (k)$ whose components $\pi _j (k)$ represent the
probability of being infected by a seeder belonging to the cluster
$j$. If $k$ is reached each of the $n_k$ times by invasion paths rooted in
premises belonging to the same cluster $m$, the vector has components
$\pi_m=1$ and $\pi_{j\neq m}=0$. On the contrary, for a node $k$
infected by epidemics originated in farms belonging to a different
cluster each of the $n_k$ times, the vector elements assume the values $\pi_j=1/n_k$.  
In the case an epidemic is detected at node $k$ by the surveillance system, 
the vector $\vec{\pi} (k)$ encodes valuable information restricting the possible set of initial conditions. In particular, it is possible
to define an uncertainty $\xi(k)$ in the identification of the seeding cluster,
by using an entropy-like function defined as $\xi(k)=-(\log n_k)^{-1}
\sum_j \pi_j \log \pi_j$. In the examples above, $\xi(k)=0$
when $k$ is always infected by the same cluster, and $\xi(k)=1$
if $k$ is infected each time by a different cluster. The normalization 
$log(n_k)$ is chosen because it represents the most homogeneous situation 
given that $n_k$ is always smaller than the total number of clusters.
An alternative normalization factor has also been tested in the ESM.

%*******************************************************************************************************
\section{RESULTS}

%************************************************
\subsection{Dynamical properties of cattle movement network}
The dynamical network of bovines displacements
exhibits complex features both in the structure of the various static
snapshots~\cite{Christley:2005,Bigras:2006,Natale:2009,Volkova:2010,Rautureau:2010}, 
and in the temporal fluctuations of links and nodes~\cite{Bajardi:2011}. 
In particular, the links lifetime and the number of displaced bovines are not characterized by a 
well-defined time-scale, and the centrality measures commonly used 
in the context of static networks appear unable to identify the most important 
nodes of the network~\cite{Natale:2011,Bajardi:2011,Kostakos:2009}.
An aggregated view of the system over a  temporal window $\Delta t$
 yields indeed a ranking of the importance of premises that 
may not reflect their properties at  different moments of the system evolution, or at other aggregation timescales (see Figure~\ref{fig:dataset}b) ~\cite{Bajardi:2011}. 

Such fluctuations 
may strongly affect spreading processes, as premises that are poorly 
connected on a given day (or week/month) may become largely connected 
on the next day (respectively, week/month) and vice versa.
Their impact on the efficacy of intervention measures is clearly important. 
Figure \ref{fig:dataset}c shows the efficacy in the reduction of the maximum possible epidemic size,
indicated by the number of premises in the giant connected component (GC), when quarantine measures are adopted that are based on the movements knowledge at a given time only. More specifically, premises are removed from a network in decreasing order of the number of connections, however this information is measured only on the $3^{rd}$
month and applied to the $3^{rd}$ and $4^{th}$ monthly networks.
While such a targeted removal is effective in rapidly
decreasing the size of the largest connected component in the network
of the $3^{rd}$ month, it is not at all effective for the network of
the successive month. This highlights how using past information might
result in ineffective containment strategies through premises isolation for such a highly varying
temporal network, and that
the characterization of the spreading properties of premises 
cannot be assessed from a topological static
point of view: 
the full dynamical nature of the trade system and of the epidemic propagating on it has to be taken into account.

%************************************************
\subsection{Epidemic profiles and dependence on the initial conditions}
\begin{figure}[H]
\centerline{\includegraphics[width=.64\textwidth]{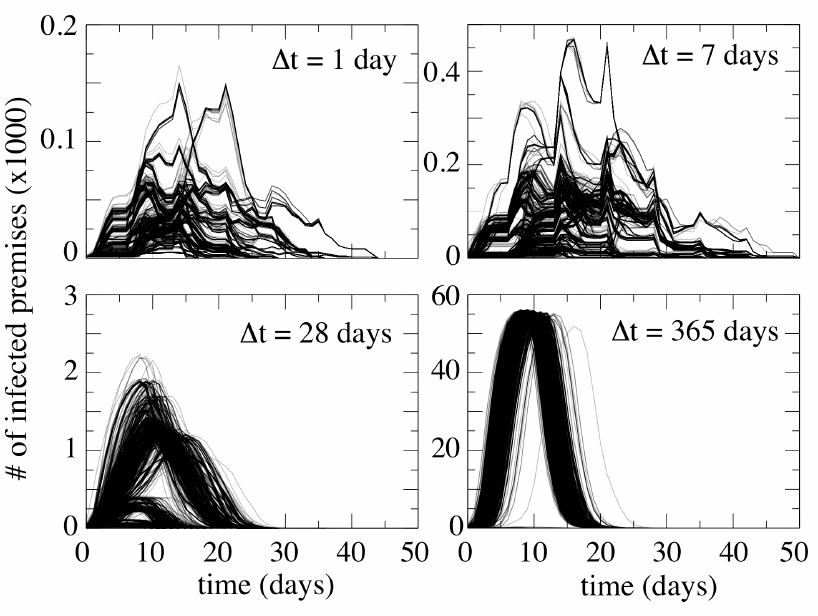}}
\caption{Number of infected premises as a function of time for different
  aggregating time windows $\Delta t$. Each curve represents the
  profile of an epidemic starting on January $1^{st}$ from a given
  seed.}
\label{prevalence_dt}
\end{figure}

We first explore the role of the aggregation timescale  $\Delta t$ of the dynamical network
on the disease propagation, by analyzing the spreading patterns resulting from outbreaks starting at 
each $x_0$ of the $\sim 1.7\cdot 10^5$ premises on the seeding date $t_0=$ January $1^{st}$,
assuming an infectious period $\mu^{-1}=7$ days.
The simulated epidemics dramatically depend on the aggregation timescale, as shown in
Figure~\ref{prevalence_dt} for daily, weekly, monthly and yearly networks.

The spreading becomes faster and reaches a larger proportion of the
nodes~\cite{Vernon:2009} with increasing $\Delta t$, as expected since the temporal
aggregation allows propagation paths that would otherwise be prevented by causality.
Most importantly, the epidemic profiles  show for short $\Delta t$ an intrinsic variability
as a function of the initial conditions of the outbreak, with
multiple peaks and strong differences in peak times for different
initial conditions. The  aggregation on large $\Delta t$ values leads  to a loss of the 
network intrinsic variability and therefore to a smaller impact of the seeds on the epidemic profiles.
The large temporal fluctuations describing    the premises' activity~\cite{Bajardi:2011}
does not allow the identification of an upper bound of the timescale $\Delta t$ that could be a good approximation to the system description, since any given infectious period $\mu^{-1}$ would have a non-negligible interplay with a broad set of timescales that are part of the full spectrum of timescales of the dynamical system.
Therefore, in order to realistically account
for the impact of the seeding on the spread of epidemics on the
dynamical network, in the following we focus on the finest temporal
scale, $\Delta t=1$ day, for the description of the bovines mobility
in the epidemic simulations.

%************************************************
\subsection{Similar spreaders and seeds' cluster emergence}

By fixing $\Delta t=1$ day, we explored the results of the epidemic
simulations starting from all possible geographical initial conditions
corresponding to $t_0=$ January $1^{st}$. We calculated the overlap
values among all possible pairs of initial conditions and filtered the
ICSN by applying a threshold value for the overlap equal to 0.8. The network separates into several connected components, leading to a natural
emergence of clusters of  initial conditions.  These represent sets of nodes that, if 
at the origin of an outbreak, would lead to similar invasion paths. Clusters are organized in a 
hierarchy depending on the value of $\Theta_{th}$, and it is interesting to note that, given the
distribution of similarity values obtained, even large enough values of $\Theta_{th}$ lead to the
emergence of a number of non-trivial clusters of initial conditions,
i.e. different from simply isolated nodes.
The distribution of the sizes of the clusters is shown in the ESM, along with a sensitivity analysis
  on the value of $\Theta_{th}$.

\begin{figure}[H]	
\begin{center}
\vspace{-0.7cm}
\centerline{\includegraphics[width=\textwidth]{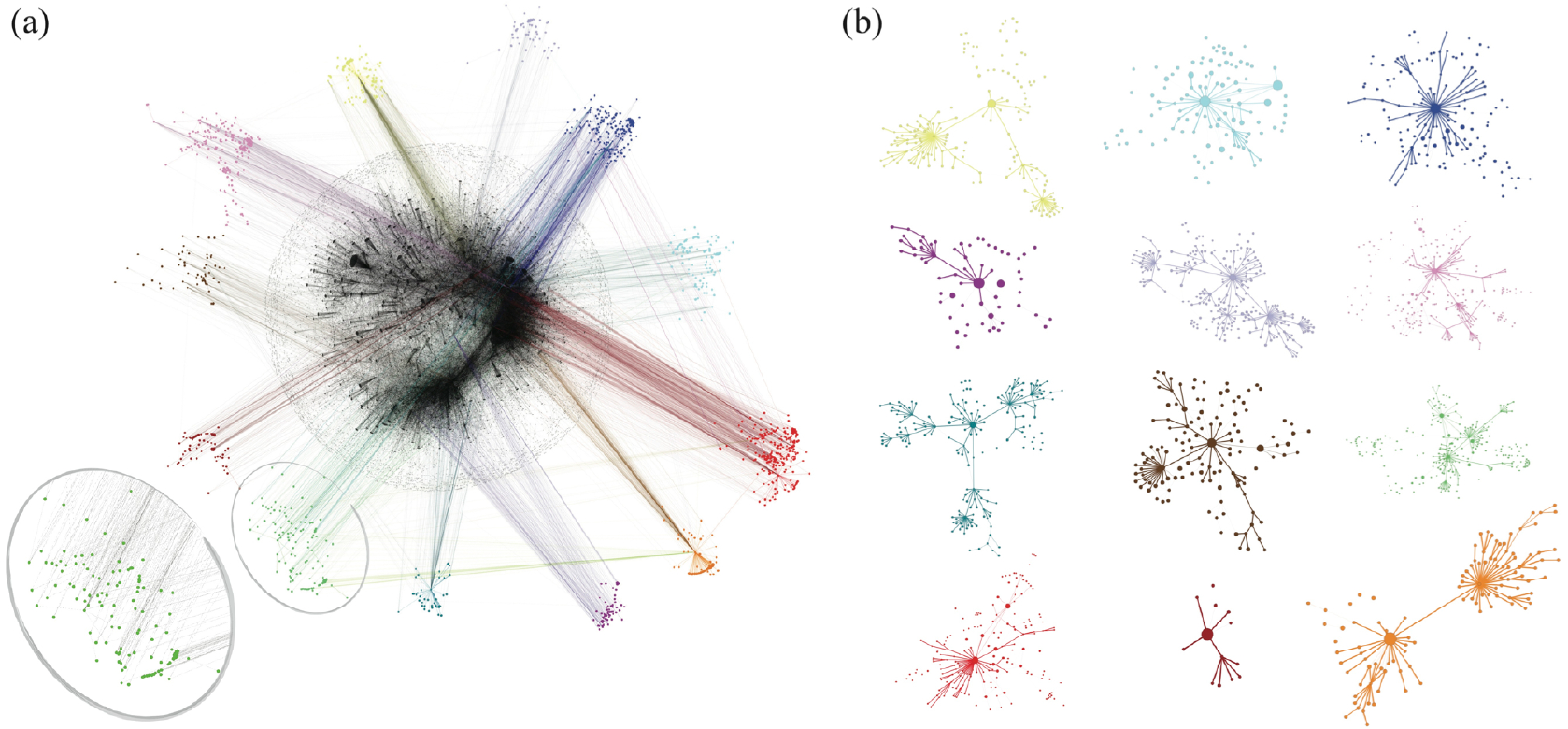}}
\caption{Topological representation of clusters and invasion paths
  corresponding to $t_0=$ January $1^{st}$. (a) The nodes belonging to
  each cluster are represented in the network
  of bovines displacement aggregated over the whole spreading period
  (grey network). The zoomed frame shows the absence of community-like
  structures or chain-like motifs.  (b) Each network represents the
  union of all invasion paths starting from the nodes of a given
  cluster.  The initial conditions are not
  shown for the sake of visualization; the link thickness is proportional to
  the number of invasion paths propagating along that connection and
  the size of the nodes is proportional to the number of incoming
  invasion paths.  Different topological structures of the invasion
  paths are found for different seed clusters. 
  All nodes belonging to a given cluster are shown with
the same color, and the same color is used in both panels for each
cluster.
 }
\label{italy_clust}
\end{center}
\end{figure}

In Figure~\ref{italy_clust} we show the $12$ largest clusters
 along with the displacement network
aggregated over the entire spreading period.
  Some important characteristics of
the clusters emerge clearly. First, the nodes of a given cluster
 are not tightly connected in the aggregated
displacement network. In addition, there
is a lack of chains of infections: the nodes in the clusters are not
trivially connected to each other by links that bring the disease from
one node to the next. A direct analysis of the aggregated displacement
network, based for instance on the search of communities or chain-like
structures, would therefore not be able to detect the similarity of their
spreading properties, even if performed using different aggregation timescales $\Delta t$.

\begin{figure}[p]	
\begin{center}
\vspace{-0.7cm}
\centerline{\includegraphics[width=0.45\textwidth]{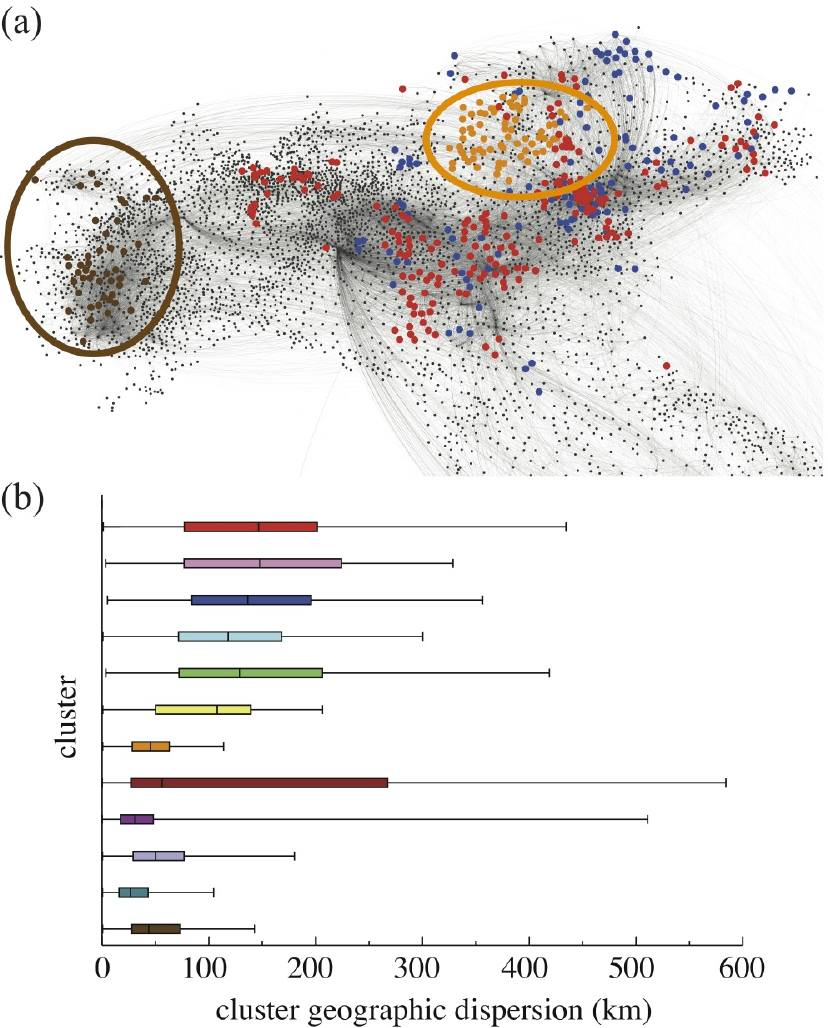}}
\caption{Geographical characterizations of clusters corresponding to
  $t_0=$ January $1^{st}$. (a) As a paradigmatic example, $4$ clusters are shown in
  different colors on the georeferenced network of bovines
  displacement aggregated over the whole spreading period ($35$
  days). The same color
  code of Figure~\ref{italy_clust} is used. The ellipses highlight the most compact
  clusters in terms of geographical dispersion. In the same area different clusters may coexist, moreover, some of them can be rather dispersed, as shown by the clusters colored in red and in blue.  (b) Cluster geographical
  dispersion, calculated as the distance between each pair of nodes
  belonging to the same cluster (identified by the color).  }
\label{italy_clust2}
\end{center}
\end{figure}

The spatial analysis of the
georeferenced representation of the clusters (where each node is
assigned the location of the corresponding municipality) shows moreover that,
although some clusters are formed by nodes which are geographically
rather close, most clusters are dispersed, with a distribution of
distances between nodes spanning several hundreds of kilometers (Figure~\ref{italy_clust2}). Clusters can also
geographically overlap and do not have mutually separated geographical
boundaries.
Therefore, the geographical proximity of two nodes does
not necessarily imply that they will lead to similar invasion paths. 

\begin{figure}[H]	
\begin{center}
\centerline{\includegraphics[width=0.44\textwidth]{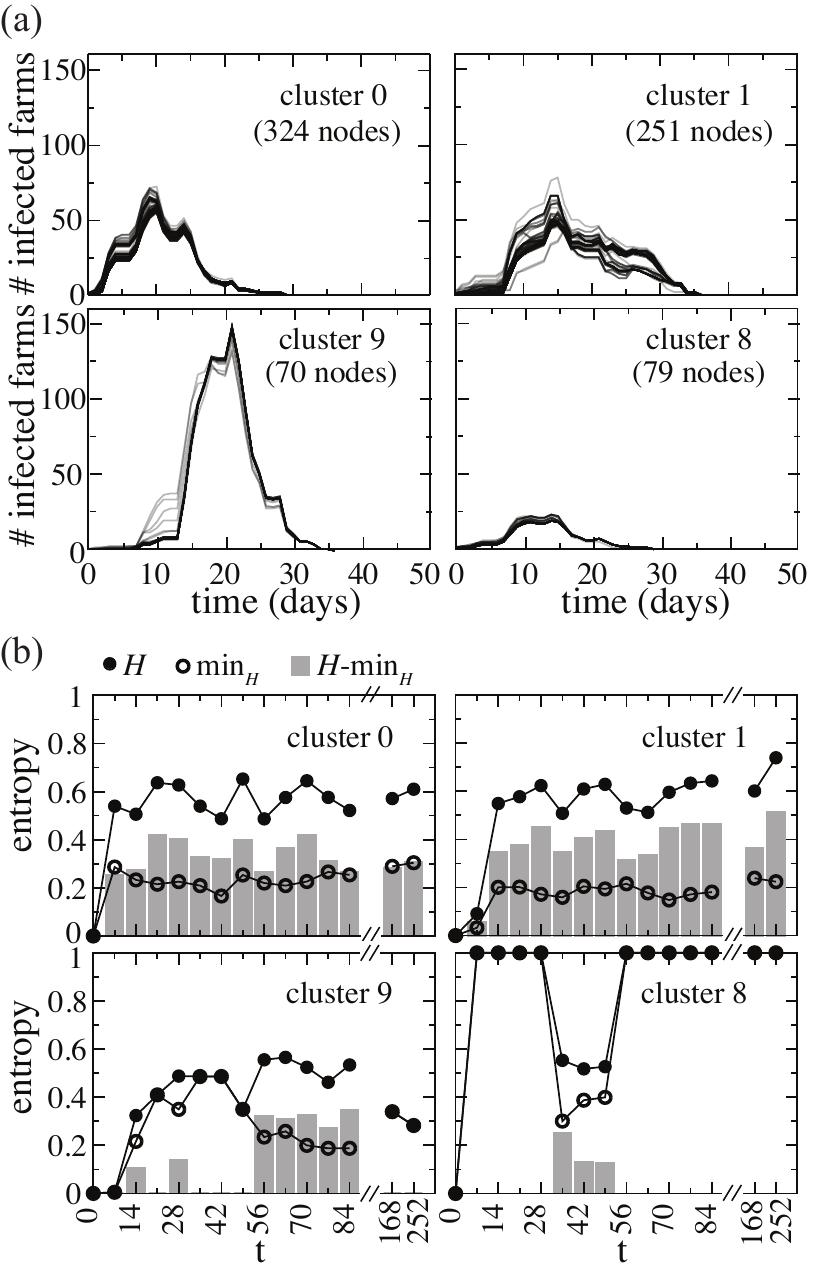}}
\vspace{-0.3cm}
\caption{Seeds' clusters characterization. (a) Number of infected farms as a
  function of time. Each panel reports the profiles of the 
  epidemics starting from initial conditions belonging to the same
  cluster. Four clusters of different sizes are shown as examples. (b)
  Entropy $H$ of the partition into clusters at $t_0=$ January $1^{st}$ as a function of time
  $t=t_0+7w$ with $w=1, 2, 3,...$, for the same clusters as in (a). The
  difference $H-\min_H$ (grey bars) represents the robustness of the
  cluster (the smaller the difference and the more robust is the
  cluster), given that only part of the nodes may survive in the partition at time $t$. 
  Four typical behaviors can be characterized, as explained in the main text, each reported
  by an example in the figure.}
  \label{prevalence_entropy_cluster}
\end{center}
\end{figure}

Overall, neither the structural nor the geographical analysis of the
dynamical network of displacements would be able to reveal the
existence and composition of groups of nodes leading to similar
spreading patterns, and a detailed analysis of the dynamical process
 is needed.  Interestingly, the mixed shapes observed in the
 profiles of Figure~\ref{prevalence_dt} are automatically
classified into a set of specific and well-defined  behaviors
by considering initial conditions belonging to the same cluster, as
shown in Figure~\ref{prevalence_entropy_cluster}a.  Grounded in the comparison of the
infected nodes and disregarding the explicit links of transmission, the
clustering method is able to group the spreading histories into
similar patterns characterized by the same timing and size. An alternative version
of the clustering method based on the overlap of the full invasion paths leads to a 
similar partition, despite the fact it relies on a much larger amount of information (see the ESM).
Similar findings are also obtained  with a stochastic infection
dynamics, as reported in the ESM.

%*************************************************************
\subsection{Longitudinal stability of the seeds' clusters}

 Given the strong
variability of the network's properties on all timescales~\cite{Bajardi:2011},
partitions obtained for spreading processes starting at different
times could substantially differ. 
In order to investigate this aspect, we compare the partition obtained at time $t_0$
with the one obtained at time $t>t_0$ by means of the entropy function $H$ defined in the Materials and
Methods section. $H$ measures the level of fragmentation of the cluster partition in time,
with small values indicating a large stability and values close to 1 indicating the disruption
of the original partition. The lower bound (min$_H$) represents the most stable configuration
and takes into account the possible disappearance of nodes from the partition at time $t$.
We present in Figure~\ref{prevalence_entropy_cluster}b the results corresponding to
$t_0=$ January $1^{st}$ and $t=t_0+7w$ with $w=1, 2, 3,...$,
i.e. successive times separated by $w$ weeks from $t_0$. 
The cluster temporal stability can roughly be
classified into four main behaviors, shown through four examples: {\it i}) a substantial
fraction of the nodes of the cluster disappears already for $w=1$
($\min_{H}\neq 0$), and small groups of nodes are redistributed in other
clusters (small  $H-\min_H$), quite stable in time 
(cluster 0); {\it ii}) high stability at $w=1$, followed by a similar behavior (cluster 1); 
{\it iii}) high stability at $w=1$, followed
by a robust preservation of the partition for several weeks (cluster 9); 
{\it iv}) a very unstable behavior, as
the cluster's nodes disappear almost completely from the
partition at time $t$ (very high $\min_{H}$, cluster 8).
The most robust behavior in time
(shown by the example of cluster 9), was found for 2 clusters out
of the 20 largest clusters considered for $t_0=$ January $1^{st}$.
Interestingly, it turns out that the size of a cluster is not
correlated with its stability, as shown in the ESM where the stability
of all clusters is investigated, along with additional measures
of stability and a sensitivity on the C values considered.

%*****************************
\subsection{Disease sentinels}

The success of control and mitigation measures critically depends on the ability
to rapidly detect an outbreak and identify its source. Ideally, a timely detection of the
origin of the disease would allow a targeted strategy  to isolate
the infected premises and contain the propagation further. Longer delays between the start of the outbreak and its
detection mean larger numbers of infected farms, a more difficult
identification of the starting point of the spreading, and therefore 
of the propagation paths, 
overall leading to increasing difficulties in preventing
further spread and to increasingly expensive containment measures.  
The high temporal variability and the complex nature of the network of
displacements makes the identification of the possible origin of the outbreak, 
following the detection of an infected node, a particularly difficult task. 
This has to be factored in with partial or missing knowledge on the
epidemic situation due to under-reporting and/or the presence of a silent spread
phase that would delay the first detection of the outbreak while propagation occurs.
The heterogeneous nature of the network allows however  the identification
of clusters of seeds leading to similar invasion paths, that may be used to enhance surveillance and help the inference of the origin of a disease, 
once an epidemic  unfolds on the network.

\begin{figure}[H]
\begin{center}
\centerline{\includegraphics[width=0.44\textwidth]{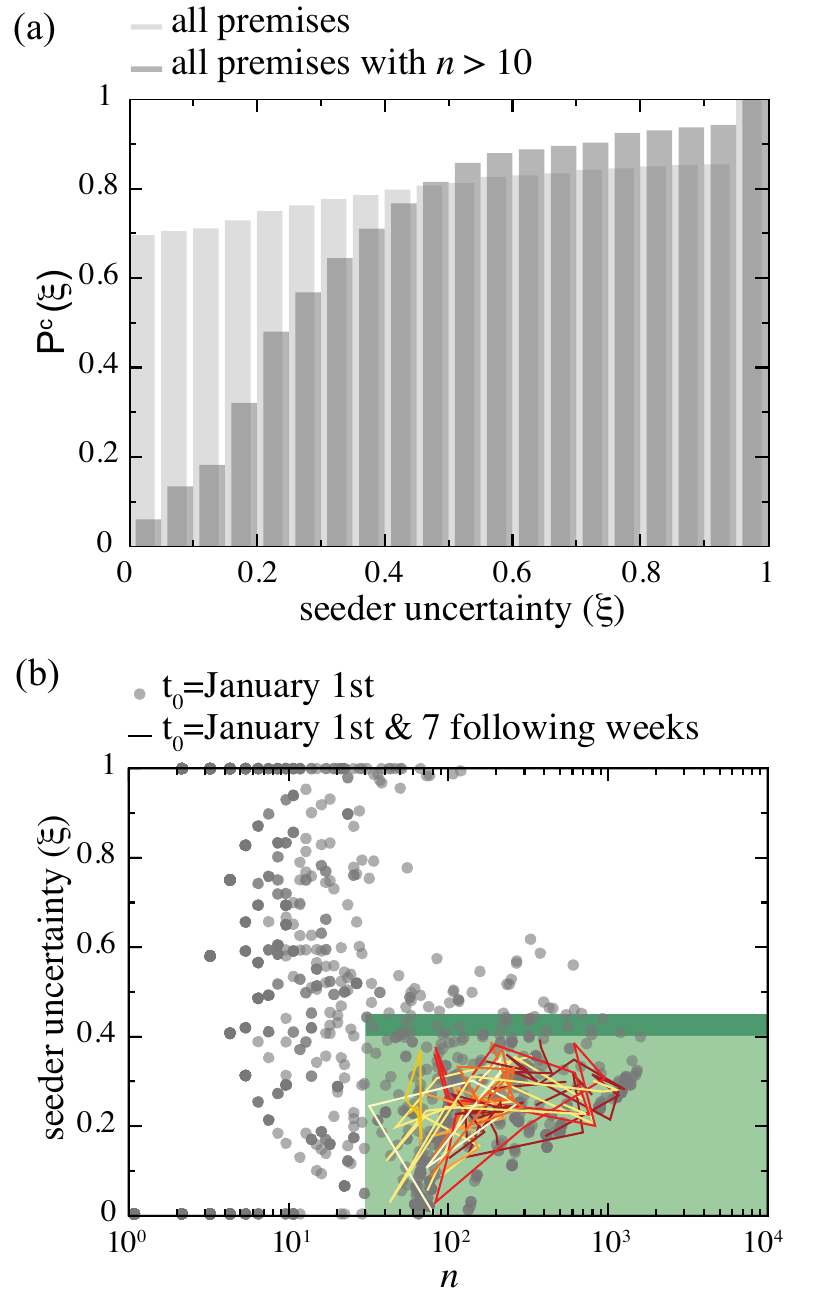}}
\vspace{-0.1cm}
\caption{Sentinel premises.
  (a) Cumulative probability distribution of the uncertainty $\xi$ of a given premises in the
  identification of the seeding cluster.
  Slaughterhouses are discarded from the
  analysis, as they cannot spread the disease further to other farms. (b) For a set of initial
  conditions $(x_0,t_0)$, with $t_0=$ January $1^{st}$, each infected farm is represented by a dot in the
  $n-\xi$ phase space, with $n$ being the number of times the farm is
  reached by an infection. Sentinel nodes are defined as the farms
  that are often reached by epidemics (i.e. $n>n_s$) and have
  a low degree of uncertainty in the identification of the seeding
  cluster that led to the outbreak (i.e. $\xi<\xi_s$). The plot 
  shows the trajectories in the $n-\xi$ phase space of the $15$ sentinels obtained
  by imposing $n_s=30$ and $\xi_s=0.4$, for eight
  consecutive weeks starting from January $1^{st}$. }
\label{sentinel}
\end{center}
\end{figure}

Based on the cluster partition ${\cal P}(t_0)$ obtained from the
epidemic simulations starting at time $t_0$ from all possible initial
conditions, we calculate the uncertainty $\xi$ of all premises
infected by the epidemic in identifying the seed cluster originating
the outbreak.  Figure~\ref{sentinel}a shows the cumulative
distribution of the uncertainty $\xi$. The number of times $n_k$ that
a holding is infected may strongly vary from one holding to the next;
in particular, many nodes are in fact infected just once ($n_k=1$),
yielding trivially high $\xi(k)$ values. We thus focus on premises
that have been infected at least $10$ times. Interestingly, even with
this restriction, the seeder uncertainty is less than $40\%$ for
almost $70\%$ of the infected nodes, meaning that most premises
reached by the infection are able to provide valuable insights about
the origin of the disease in terms of the identification of the
cluster from which the spreading originated. As a result, information
about the invasion paths and the epidemic timing is also obtained,
following the findings of Figure~\ref{prevalence_entropy_cluster}a.

The uncertainty $\xi(k)$ on the identification of the cluster of initial conditions
infecting the node $k$ and the number of times $n_k$ the node $k$ is reached by the epidemic
clearly depend on the time $t_0$ of the start of the epidemic. In the following, we
explore the variation of these two quantities for all nodes of the network when
we consider epidemics starting at time $t_0=$ January $1^{st}$ $+7w$ with 
$w=0, 1, 2, 3,..., 8$, i.e., spanning an 8-weeks interval from January $1^{st}$. 
In Figure~\ref{sentinel}b we represent each farm $k$ as a point  with coordinates $(n_k,\xi(k))$ in the
$n-\xi$ phase space, for $t_0$= January $1^{st}$. As $t_0$ changes,
a variety of different behaviors is obtained, as expected given the large variability of the network.
Large fluctuations of the number of times a node is infected are observed, as a node
with a large $n_k$ (i.e., often reached by the disease) for
an initial time $t_0$ may be rarely reached if the outbreak
starts later, given the change in the network of displacements, or may even disappear
from the plot if it is not infected for a given explored initial time (i.e., it has $n_k=0$).
Similarly, also the values of the uncertainty in the identification of the seeding cluster can strongly
fluctuate. From the surveillance perspective, we are interested in the nodes that 
are infected a large number of times (i.e., are likely reached by the epidemic, given any temporal and geographical
initial conditions) and for which we have a low uncertainty in the identification of the seeding cluster,
providing important insights into the previous and future spreading patterns. 
We define these premises as {\em sentinel nodes} by imposing that they are infected
at least $n_s$ times and are characterized by an uncertainty at most equal
to $\xi_s$ for all initial conditions. Their trajectories in the $n-\xi$ phase space
for varying $t_0$ are shown in Figure~\ref{sentinel}b for
$n_s=30$ and $\xi_s=0.4$.
The choice of the $(n_s,\xi_s)$ threshold values depends on the resources
available to monitor these sentinels: smaller $n_s$ and larger
$\xi_s$ lead to a larger number of sentinels. In Table 1 of the ESM, we
report the number of sentinels for different $(n_s,\xi_s)$ values.
It is also possible to be less conservative
and enlarge the group of possible sentinels for an efficient detection
of an infectious disease by including farms with discontinuous
trajectories that may have $n_k=0$ for one value of the starting time but
have $n_k \ge n_s$ and $\xi_k \le \xi_s$ for the other starting
times. By relaxing these constraints, it is possible to build a hierarchy of disease
sentinels with different levels of reliability, and specific to the available surveillance resources.
In the ESM we also tested an alternative definition of the  entropy function 
showing that it does not alter the results.

\begin{figure}[p]
\begin{center}
\centerline{\includegraphics[width=\textwidth]{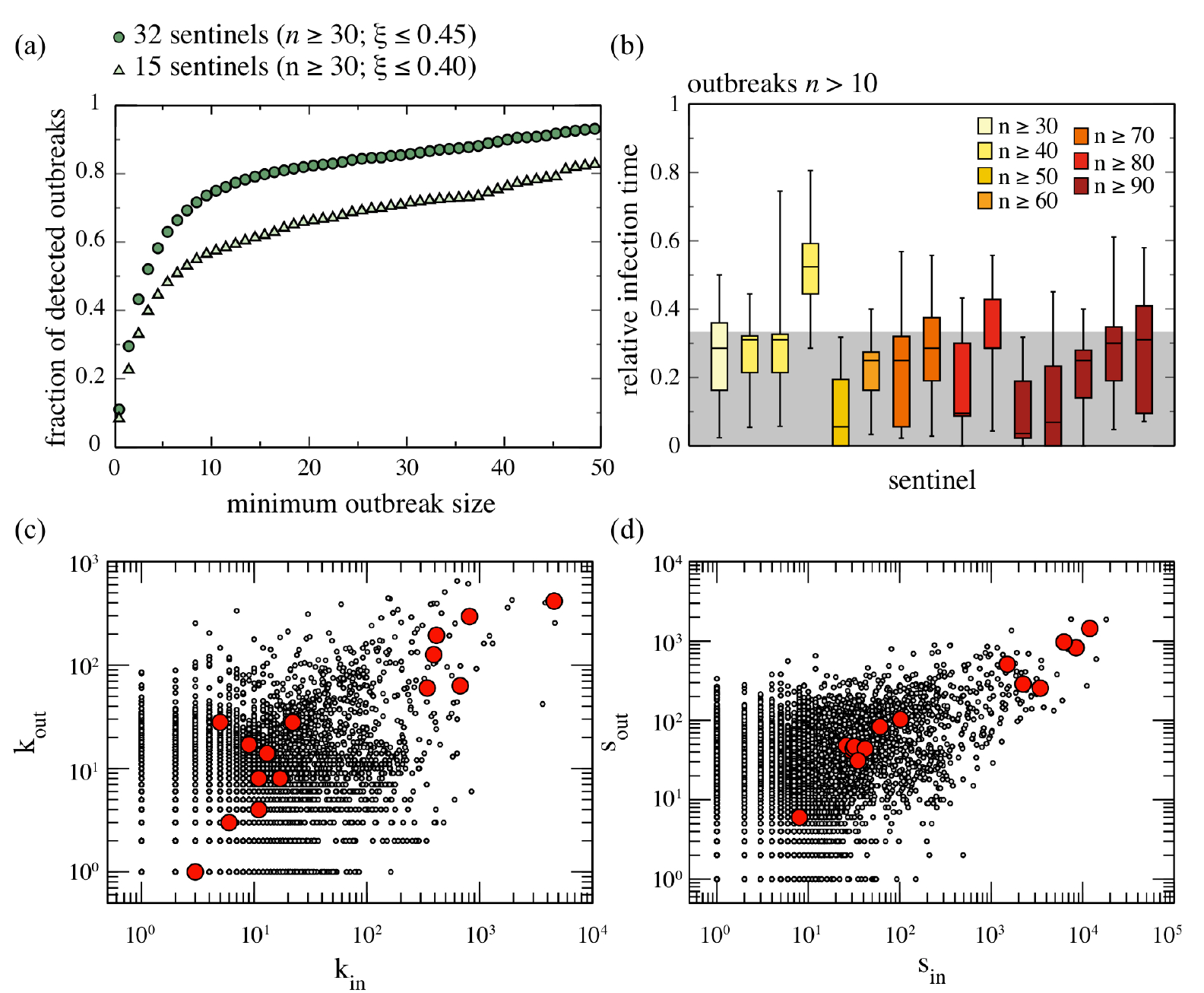}}
\vspace{-0.1cm}
\caption{Properties of the surveillance system based on sentinel premises.
  (a)  Fraction of 
  outbreaks detected by the sentinels as a function of the minimum outbreak size of the
  epidemic, for two sets of sentinels (of $15$ and $32$ sentinels), corresponding 
  to ($n_s=30$, $\xi_s=0.4$) and ($n_s=30$, $\xi_s=0.45$), respectively. 
 (b) Boxplot of the time of infection of
  the $15$ sentinels relative to the full duration of the outbreak,
  considering the detected outbreaks with final size larger than
  $10$. Each box is colored according to the number of times $n$ that the
  sentinel has been infected and a grey shaded area indicates
    $33\%$ of the relative infection time.
    (c) Topological properties of sentinel nodes (red dots), compared to the other nodes (smaller black dots). All premises in the system (except slaughterhouses) are represented in the plane of the number of in-connection vs. the number of out-connections per premises. Sentinels may be characterized by either small or large number of in/out connections.
    (d) As in (c), but showing the fluxes properties in the plane of the number of batches moved in and out of each premises. Even in this case, sentinels may assume small to large values in the parameter space.
    }
\label{sentinel2}
\end{center}
\end{figure}

The interest of the definition of sentinel nodes in the perspective of
a surveillance system is quantified further in
Figure~\ref{sentinel2}a.  Given a set of sentinels, we measure the
fraction of detected outbreaks as a function of the outbreak final
size, where an outbreak is considered detected if it infects at least
a sentinel farm.  Figure~\ref{sentinel2}a shows that sentinels are not
good indicators for the presence of small outbreaks (i.e.,
corresponding to sizes smaller than 5-10 infected farms), however a
surveillance system based on only $15$ sentinel nodes (out of a total
number of more than $170,000$ premises) would detect more than $55\%$
of the outbreaks with final size larger than $10$ and, if the number
of sentinels is increased to $32$, the fraction of outbreaks detected
would be more than $75\%$. Finally, it is also important to consider
that the information provided by the sentinel farms is meaningful as
long as the detection occurs rather early in the outbreak evolution.
We evaluated the rapidity of the detection by plotting the infection
time of each of the $15$ sentinel farms (obtained with $n_s=30$ and
$\xi_s=0.4$) relative to the full outbreak duration, for outbreaks
with size larger than $10$ (Figure~\ref{sentinel2}b). Interestingly, almost all sentinels are able
to detect most outbreaks within the first third of the outbreak
duration. 
Similar results are also valid for the stochastic case, as reported in the ESM.

It is finally interesting to note how sentinel nodes cannot be identified through 
geographical, topological or fluxes analyses only. Figure 8c-d shows the properties
of the sentinel nodes in terms of the number of in- and out-connections, and of the number of
 batches moved in and out of the premises, highlighting how sentinels do not share  similar
 properties and span largely fluctuating values in the parameter space.

%*******************************************************************************************************
 \section{DISCUSSION}

The full knowledge of the livestock movements at
 a daily resolution makes it possible to investigate in detail
the  spreading patterns of livestock emerging diseases. Through simulations
on the fully dynamic network, where daily bovine movements are explicitly
captured, we have studied the role of the initial conditions in shaping
the propagation process. Clusters of seeds emerge that lead to similar spreading
patterns in terms of infected premises, and are also characterized by similar 
epidemic profiles and peak times. These clusters cannot be identified 
from purely structural or geographical considerations. The proposed clustering method 
can be used in order to optimize surveillance systems and 
define rapid and efficient containment strategies, targeting
farms that are at high risk of being infected and further spread the disease.
Although the displacement network is characterized by a large temporal variability,
intrinsically altering the centrality role of nodes from a given observation time to another, 
it is possible to identify sentinel nodes representing premises which are often 
reached by the disease and, when detected as infected, are able to provide valuable information
on the seeding farms of the outbreak and thus on the likely spreading path, allowing to design targeted intervention
  strategies.
 A hierarchical classification of sentinels can be provided by tuning the constraints imposed
for their definition, leading to different levels of surveillance. 
 Remarkably, the bare
   knowledge of  the animal movements would not be enough to estimate
   the origin of a disease, once detected, as the outbreak results
   from the complex interplay of the dynamical network and the
    disease dynamics.
   On the other hand, this interplay leads to the emergence of a very
small number of sentinels, with respect to the total number of premises present in the system,
that may be efficiently used for disease prevention and control.

Applications to specific diseases,
where the timescale of the epidemic is set by the parameters describing the disease etiology,
can be performed to tune this framework to particular cases.
These findings clearly depend on the 
full knowledge of the displacement dataset, and can thus be obtained as {\em a priori} information
during a non-emergency period to help orienting control strategies, as commonly done
with the static analysis of the contact network structure, 
strengthening the importance of such data collection.
 The ability to make useful predictions for current and future
livestock movements patterns 
depend on the level of similarity across different years of data. The
analysis of 
successive years of movements data, uncovering possible recurrent patterns and seasonal behaviors, may thus contribute
to make this framework a general tool to be used in real-time emergencies.\\

This work was partially funded by the European Research Council Ideas contract no. ERC-2007-Stg204863 (EPIFOR) to VC and PB;  
the EC-FET contract no. 233847 (DYNANETS) to VC; the MSRCTE0108 project funded by the Italian Ministry of Health to LS and VC.

\end{document}